\documentclass[12pt]{iopart}

\expandafter\let\csname equation*\endcsname\relax
\expandafter\let\csname endequation*\endcsname\relax

\usepackage{amsmath,amssymb,revsymb}
\usepackage{comment}
\usepackage{graphicx}

\begin{document}

\title{Spectroscopy of light atoms and bounds on physics beyond the standard model}

\author{R. M. Potvliege}

\address{Department of Physics, Durham University, South Road, Durham DH1 3LE, UK}
\ead{r.m.potvliege@durham.ac.uk}

\begin{abstract}
Newly calculated bounds on the strength of the coupling of an electron to a proton or a neutron by a fifth force are presented. These results are
derived from the high precision spectroscopic data currently available for hydrogen,
deuterium, helium-3 and helium-4. They do not depend on specific assumptions on how the interaction would couple to a deuteron compared to a proton or would couple to an $\alpha$ particle compared to a helion. They depend on its coupling to a muon, but not in a significant way for carrier masses below 100~keV if one assumes that the strength of the interaction with a muon would be of a similar order of magnitude as the strength of the interaction with an electron in that mass region.
\end{abstract}

\maketitle

\section{Introduction}

A growing number of atomic systems are relevant in the ongoing search for a new physics interaction in view of the very high level of precision achieved in their spectroscopy. Two main approaches have been considered on this front in regard to the possible existence of a fifth force. One is to search for departures from the predictions of the standard model for differences in transition frequencies between different isotopes of a same species. This approach is based on the analysis of what is called King plots nonlinearities \cite{Berengut2018}. It has been recently applied to ytterbium atoms, ytterbium ions and calcium ions (\cite{Hur2022,Door2024,Wilzewski2024} and references therein). The other is to search for departures from the predictions of the standard model in one- or two-electron systems whose transition frequencies can be both measured and calculated to a suitably high precision \cite{Jaeckel2010,Karshenboim2010b,Brax2011}.
This second approach has been explored in some detail in \cite{Delaunay2017}, in particular in regard to the prospects offered by transitions in hydrogen, deuterium, helium-3 and helium-4 for setting bounds on the strength of a fifth force, and also in our more recent work on bounds based on hydrogen and deuterium spectroscopy \cite{Jones2020,Potvliege2023}. It has been extended to a broader variety of atomic systems and experimental data for specific new physics models in \cite{Delaunay2023}.

We revisit and continue some of these earlier investigations in the present article, in the light of recent experimental and theoretical advances in the spectroscopy of light elements and their muonic counterparts \cite{Pachucki2018,Krauth2021,Lensky2022,Pachucki2024,Pachucki2024b,vanderWerf2023,Schuhmann2023,LiMuli2024,Qi2024}. 
Specifically, we consider the possibility that a new physics interaction impart a potential energy $V_{\rm NP}(r)$ to an electron or muon located at a distance $r$ from a hydrogen or helium nucleus, with
\begin{equation}
V_{\rm NP}(r) = (-1)^{s+1}\frac{g_l g_N}{4 \pi}~\frac{1}{r}~\exp(-m_{X_0} r)
\label{eq:yukpot2}
\end{equation}
in natural units. Here $m_{X_0}$ is the mass of the new physics boson mediating the interaction, $s$ is the spin of this boson, and $g_l$ and $g_N$ are two dimensionless constants ($g_l \equiv g_e$ for an electron, $g_l \equiv g_\mu$ for a muon, $g_N \equiv g_p$ for hydrogen-1, $g_N \equiv g_d$ for deuterium, $g_N \equiv g_{h}$ for helium-3 and $g_N \equiv g_{\alpha}$ for helium-4).
A large class of new physics models give rise to such a contribution to the Hamiltonian. Given the form of $V_{\rm NP}(r)$, the corresponding new physics interaction is attractive when $(-1)^{s+1}{g_l g_N} < 0$ and repulsive when $(-1)^{s+1}{g_l g_N} > 0$. 

As is commonly done in this context, we will assume that
\begin{equation}
g_n = g_d - g_p = g_\alpha - g_h,
\label{eq:gn}
\end{equation}
where $g_n$ is the coupling constant for a neutron.
Our main results are new upper bounds on the products $g_eg_p$ and $g_eg_n$. While we make use of the nuclear rms charge radii derived from Lamb shift measurements on the muonic species, we take into account the possibility that a new physics interaction might need to be taken into account in the calculation of these charge radii. We do not use scattering data in view of the difficulties with deriving charge radii from these results \cite{Khabarova2021,Hiyama2024}.

The proton rms charge radius ($r_p$), the deuteron rms charge radius ($r_d$) and the Rydberg constant ($\cal R$) are co-determined from a set of experimental and theoretical results including, in particular, high precision spectroscopic measurements in muonic hydrogen and deuterium 
($\mu$H, $\mu$D) and in electronic hydrogen and deuterium (eH, eD) \cite{Tiesinga2021}. In principle, these data may be significantly affected by the hypothetical fifth force considered in the present work. As a consequence, setting bounds on the strength of this force involves redetermining these quantities. However, doing so is hampered by well known discrepancies and inconsistencies: discrepancies between the measurements on the muonic species and the measurements on the electronic species and inconsistencies between the latter. 
These differences result, {\it inter alia}, in a significant scatter in the values of $r_p$ derived from these measurements. The value of $r_d$ derived from measurements in muonic deuterium is also in significant tension with the values that can be derived from the spectroscopy of electronic deuterium \cite{Pohl2016}. However, there is now excellent agreement between measurements on the muonic species and measurements on the electronic species in regard to the difference $r_d^2-r_p^2$, when this difference is derived directly from the isotope shift of the 1s$_{}$~--~2s$_{}$ interval of eH and eD \cite{Pachucki2024}.

The relevant experimental results are briefly surveyed in Section~\ref{section:data}.
We calculate bounds based on hydrogen and deuterium spectroscopy in Section~\ref{section:2}, deriving them either from the isotope shift of the 1s~--~2s interval \cite{Delaunay2017} or through a global fit of the data to theoretical models \cite{Jones2020,Potvliege2023}. The former approach is extended to helium in Section~\ref{section:3}.  


\section{Current data}
\label{section:data}

\subsection{Hydrogen and deuterium spectroscopy}
\label{section:HDspectroscopy}

\begin{figure}
\begin{center}
\includegraphics[width=1.0\textwidth]{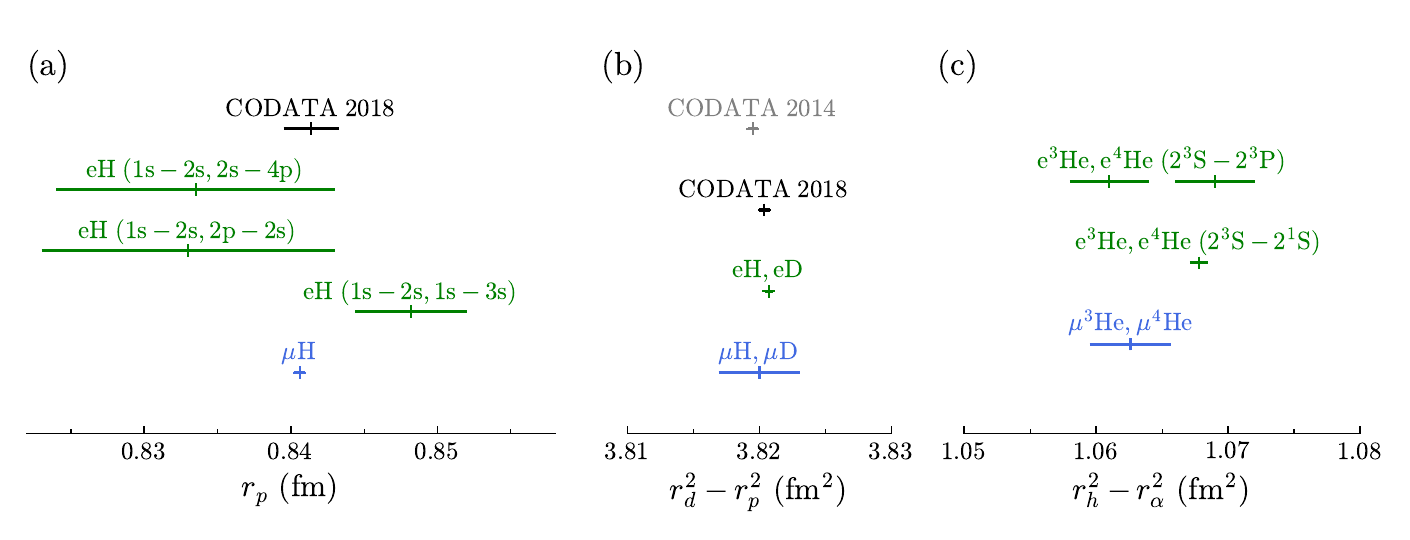}
\end{center}
\caption{(a) The proton rms charge radius, $r_p$. (b) The difference $r_d^2-r_p^2$. (c) The difference $r_h^2 - r_\alpha^2$. From top to bottom in (a): Value recommended by CODATA (2018 adjustment) \cite{Tiesinga2021}; value derived from measurements of the 1s$_{}$~--~2s$_{}$ and 2s$_{}$~--~4p$_{}$ intervals in eH \cite{Beyer2017}; value derived from measurements of the 1s$_{}$~--~2s$_{}$ interval and of the 2p$_{}$~--~2s$_{}$ Lamb shift in eH \cite{Bezginov2019}; value derived from measurements of the 1s$_{}$~--~2s$_{}$ and 1s$_{}$~--~3s$_{}$ intervals in eH \cite{Grinin2020}; value derived from measurements in muonic hydrogen \cite{Pachucki2024}.  From top to bottom in (b): Value recommended by CODATA (2014 adjustment) \cite{Mohr2016}; values recommended by CODATA (2018 adjustment) \cite{Tiesinga2021}; result derived from the isotope shift of the 1s$_{}$~--~2s$_{}$ interval in the electronic species \cite{Pachucki2018}; value derived from measurements in muonic hydrogen and muonic deuterium \cite{Pachucki2024}. From top to bottom in (c): Values derived from the isotope shift of the 2$\,{}^3$S~--~2$\,{}^3$P interval in the electronic species as measured by Shiner {\it et al}  \cite{Pachucki2017,Shiner1995} (left) or as measured by Cancio Pastor {\it et al} \cite{Pachucki2017,CancioPastor2004,CancioPastor2012} (right); value derived from the isotope shift of the 2$\,{}^3$S~--~2$\,{}^1$S interval in the electronic species \cite{Pachucki2024b,vanderWerf2023}; value derived from measurements in muonic $^3$He and muonic $^4$He \cite{Krauth2021,Schuhmann2023} as redetermined in \cite{LiMuli2024}.}
\label{fig:diffsq}
\end{figure}
Fig.~1(a) and similar figures in \cite{Brandt2022,Scheidegger2024} illustrate the current situation in regard to high precision spectroscopy of eH and $\mu$H. The measurements in $\mu$H yield a value of $r_p$ of 0.84060(39)~fm \cite{Pachucki2024,Antognini2013} (we denote this value by $r_{p,\mu{\rm H}}$
in the following). The values derived from measurements in eH have a larger uncertainty. The most precise published so far are based on the 1s$_{}$~--~3s$_{}$, 2p$_{}$~--~2s$_{}$, 2s$_{}$~--~4p$_{}$ or 2s$_{1/2}$~--~8d$_{5/2}$ intervals, in conjunction with previous measurements of the 1s$_{}$~--~2s$_{}$ interval \cite{Parthey2011,Matveev2013}. A value of 0.8482(38)~fm can be derived from the most recent measurement of the 1s$_{}$~--~3s$_{}$ interval \cite{Grinin2020}. It is in $2\,\sigma$ tension both with $r_{p,\mu{\rm H}}$ and also with the still larger value of 0.877(13)~fm derived from an independent measurement of the same interval \cite{Fleurbaey2018}. A recent measurement of the 2s$_{1/2}$~--~8d$_{5/2}$ interval yield a value of 0.8584(51)~fm, larger than and in $3.5\,\sigma$ tension with $r_{p,\mu\mathrm{H}}$ \cite{Brandt2022} and differing by $2.2\,\sigma$ from the still larger value implied by the results of a previous measurement of that interval \cite{deBeauvoir1997}. On the other hand, the values of $r_p$ based on the recent measurements of the 2p$_{}$~--~2s$_{}$ and 2s$_{}$~--~4p$_{}$ intervals, respectively 0.833(10)~fm \cite{Bezginov2019} and 0.8335(95)~fm \cite{Beyer2017}, are in good agreement with each other and with $r_{p,\mu\mathrm{H}}$.

By contrast, and as can be seen, e,g, from Fig.~1(b), the value of the difference $r_d^2-r_p^2$ derived from the measurements on the muonic species is in excellent agreement with the value derived from the isotope shift of the 1s$_{}$~--~2s$_{}$ interval in the electronic species \cite{Pachucki2024}. The experimental uncertainty on this difference is remarkably small for the electronic species, owing both to the particularly high precision with which this isotope shift was measured \cite{Parthey2010} and to the cancellation of theoretical errors in the final results \cite{Pachucki2018,Jentschura2011b}.

\subsection{$^3$He and $^4$He spectroscopy}
\label{section:Hespectroscopy}
As discussed in \cite{Delaunay2017}, setting bounds based on individual helium transition frequencies is hampered by relatively large experimental or theoretical uncertainties for most of these transitions. This issue can be avoided by using the difference of the squares of the rms charge radii of $^3$He and $^4$He instead, $r_h^2-r_\alpha^2$, for which a set of highly precise experimental results is now available --- see, e.g., Fig.~1(c).
Measurements in muonic $^3$He and muonic $^4$He have resulted in a value of 1.0636(31)~fm$^2$ for $r_h^2-r_\alpha^2$ \cite{Krauth2021,Pachucki2024,Schuhmann2023}, or 1.0626(29)~fm$^2$ as redetermined in \cite{LiMuli2024}. 
The most precise determination of $r_h^2-r_\alpha^2$ in the electronic species to date, 1.0678(7)~fm$^2$, is based on recent measurements of the  $2\,{}^3$S~--~$2\,{}^1$S interval in helium-3 \cite{vanderWerf2023} and helium-4 \cite{Rengelink2018}, combined with theory \cite{Pachucki2024b,Qi2024,Pachucki2017}; this value differs by $1.3\,\sigma$ and $1.7\,\sigma$ from those deduced from the muonic species in, respectively, \cite{Schuhmann2023} and \cite{LiMuli2024}. The result derived from a previous measurement of that interval in helium-3 \cite{vanRooij2011} also agrees with these two values but has a much larger uncertainty \cite{vanderWerf2023}. 
Three experimental values of the $2\,{}^3$S~--~$2\,{}^3$P isotope shift are currently available \cite{Pachucki2017,Shiner1995,CancioPastor2004,CancioPastor2012,Zheng2017}. Two give values of $r_h^2-r_\alpha^2$ in excellent agreement with the 
value of 1.0636(31)~fm$^2$ derived from the measurements on the muonic species but are in $2\,\sigma$ tension with each other. The third one gives a considerably different result, 1.028(2)~fm$^2$ \cite{Zheng2017}. The value derived from the $2\,{}^1$S~--~$3\,{}^1$D interval, 1.059(25)~fm$^2$ \cite{Huang2020}, is also in agreement with the muonic value but has a considerably larger uncertainty.

\section{Bounds derived from hydrogen and deuterium spectroscopy}
\label{section:2}

\subsection{General approach: Method}
\label{section:general}

Energy differences between states of electronic hydrogen or deuterium are usually expressed as transition frequencies, e.g., $\nu_{ba}$ for the energy difference between a state $b$ and a state $a$. Theoretically, these transition frequencies have the following general form within the standard model, 
\begin{equation}
    \nu_{ba}^{\rm SM}({\cal R}, r_p,r_d) = {\cal R}\,{\eta}_{ba}^{\rm g}
    + r_p^2\,{\eta}_{ba}^{\rm ps} + r_d^2\,{\eta}_{ba}^{\rm ds} + \nu_{ba}^{\rm oc},
    \label{eq:Deltath}
\end{equation}
where ${\nu}_{ba}^{\rm oc}$, ${\eta}_{ba}^{\rm g}$, ${\eta}_{ba}^{\rm ps}$, ${\eta}_{ba}^{\rm ds}$ and ${\nu}_{ba}^{\rm oc}$ are constants. These constants do not need to be calculated with highly precise values of ${\cal R}$, $r_p$ and $r_d$ in order to obtain $\nu_{ba}^{\rm SM}({\cal R}, r_p,r_d)$ with a precision matching the corresponding experimental transition frequency. The term ${\cal R}\,{\eta}_{ba}^{\rm g}$ accounts for the gross structure of the spectrum as predicted by the non-relativistic theory to leading order in the fine structure constant, the terms $r_p^2\,{\eta}_{ba}^{\rm ps}$ and $r_d^2\,{\eta}_{ba}^{\rm ds}$ account for the bulk of the dependence of $\nu_{ba}$ on the nuclear charge radii, and the term $\nu_{ba}^{\rm oc}$ accounts for all the other relevant relativistic and QED corrections. The deuteron size term $r_d^2\,{\eta}_{ba}^{\rm ds}$ is absent for transitions in hydrogen-1, and conversely the proton size term $r_p^2\,{\eta}_{ba}^{\rm ps}$ is absent for transitions in deuterium. Similar expressions relate $r_p$ and $r_d$ to the Lamb shift in muonic hydrogen and muonic deuterium. 

Given these expressions, the values of ${\cal R}$, $r_p$ and $r_d$ must be such that these theoretical energy differences match the measured intervals within experimental and theoretical errors. Namely, they must be such that
\begin{equation}
\nu_{b_ia_i}^{\rm SM}({\cal R},r_p,r_d) \doteq
    \nu_{b_ia_i}^{\rm exp}, \quad i = 1,2,3,\ldots
\end{equation}
over all the transitions considered if the possibility of a new physics interaction is ignored. (Since this set of equations is overdetermined in most cases of interest, the resulting values of ${\cal R}$, $r_p$ and $r_d$ normally need to be obtained by $\chi^2$-fitting, as the symbol $\doteq$ indicates \cite{Mohr2000}.) A hypothetical fifth force would contribute a new physics shift of $\nu_{b_ia_i}^{\rm NP}$ to the measured interval, for a transition between a state $a_i$ and a state $b_i$. If one assumes the existence of this interaction, comparing experiment to the standard model then involves finding values of ${\cal R}$, $r_p$ and $r_d$ such that
\begin{equation}
\nu_{b_ia_i}^{\rm SM}({\cal R},r_p,r_d) \doteq
    \nu_{b_ia_i}^{\rm exp} - \nu_{b_ia_i}^{\rm NP}, \quad i = 1,2,3,\ldots
\label{eq:constraints1}
\end{equation}
Requiring the existence of values of ${\cal R}$, $r_p$ and $r_d$ consistent with these equations and with the measurements in muonic hydrogen and muonic deuterium is the main constraint we use for setting bounds on the strength of this fifth force. We calculate the necessary values of ${\nu}_{ba}^{\rm oc}$, ${\eta}_{ba}^{\rm g}$, ${\eta}_{ba}^{\rm ps}$, ${\eta}_{ba}^{\rm ds}$ and ${\nu}_{ba}^{\rm oc}$ as explained in Appendix~C of \cite{Jones2020} and Appendix~B of \cite{Potvliege2023}. Like \cite{Potvliege2023}, we also follow \cite{Lensky2022} and \cite{Antognini2013} for the muonic species. We calculate the new physics shifts $\nu_{ba}^{\rm NP}$ as outlined in \ref{section:appendixA} of the present article.

We set a further constraint on the strength of this hypothetical new physics interaction by requiring that the above calculations of $r_p$ and $r_d$ result in a value of $r_d^2-r_p^2$ consistent with the value determined from the experimental isotope shift of the 1s$_{}$~--~2s$_{}$ interval in the electronic species. Specifically, we require that the experimental isotope shift of that interval matches its theoretical prediction, taking into account the possibility of a new physics contribution. As is explained in Appendix~C of \cite{Potvliege2023}, this requirement can be expressed by the inequality
\begin{equation}
|\Delta| \leq 1.96\,\sigma_{\Delta},
\end{equation}
where
\begin{align}
\Delta &= 5233.27(42)~\mbox{kHz} + \Delta\nu_{2{\rm s}1{\rm s}}^{\rm NP} -
\frac{7\alpha^4 m_e c^2}{12 h\lambdabar_{\rm C}^2} \left[\left(\frac{m_{\rm r}^{e{\rm D}}}{m_e}\right)^3 r_d^2 - \left(\frac{m_{\rm r}^{e{\rm H}}}{m_e}\right)^3 r_p^2\right],
 \label{eq:boundC}
\end{align}
and $\sigma_{\Delta}$ is the combined experimental and theoretical error on the value of $\Delta$. In this last equation,
$h$ is Planck's constant, $\alpha$ is the fine structure constant, $\lambdabar_{\rm C}$ is the reduced Compton wavelength, $m_e$ is the electron mass, $m_{\rm r}^{e{\rm H}}$ and $m_{\rm r}^{e{\rm D}}$ are the reduced masses of the respective isotopes, and $\Delta\nu_{2{\rm s}1{\rm s}}^{\rm NP}$ is the hypothetical contribution of the new physics interaction to the experimental isotope shift of the 1s$_{}$~--~2s$_{}$ interval.

To obtain bounds on the products $g_eg_p$ and $g_eg_n$, we $\chi^2$-fit the model to the data for set values of the mass $m_{X_0}$, of the ratio $g_d/g_p$ and of the ratio $g_\mu/g_e$, subject to the aforementioned constraints and to the assumption that $g_d = g_p + g_n$.
Doing so results in upper bounds on $|g_eg_p|$ and $|g_eg_n|$, namely bounds $|g_eg_p|_{\rm max}$ and $|g_eg_n|_{\rm max}$ depending both on the carrier mass and on the ratios $g_d/g_p$ and $g_\mu/g_e$.\footnote{These two quantities are not independent since $g_n = (g_d/g_p - 1)g_p$ in view of Eq.~(\ref{eq:gn}).} 
We take $|g_eg_p|_{\rm max}$ and $|g_eg_n|_{\rm max}$ to be the largest values of $|g_eg_p|$ and $|g_eg_n|$ for which $Q(\chi^2,\nu) \geq 0.05$ where $Q(\chi^2|\nu)$ is the upper tail cumulative distribution function for the relevant number of degrees of freedom, $\nu$ (the boundary value of 0.05 corresponds to a confidence level of 95\% that the data exclude the possibility that $|g_eg_p|$ and $|g_eg_n|$ are larger than the values of, respectively, $|g_eg_p|_{\rm max}$ and $|g_eg_n|_{\rm max}$ obtained by the fitting procedure). For each value of $m_{X_0}$ and of $g_\mu/g_e$, we then take the absolute upper bound on $|g_eg_p|$ to be the highest value of $|g_eg_p|_{\rm max}$ over the range $-\infty < g_d/g_p < \infty$, and similarly for the absolute upper bound on $|g_eg_n|$. We found that varying $g_d/g_p$ between $-1$ and $3$ was sufficient for finding these absolute maxima for most values of $m_{X_0}$.

\subsection{General approach: Results for $g_\mu = g_e$}
\label{section:results}

\begin{figure}
\begin{center}
\includegraphics[width=0.9\textwidth]{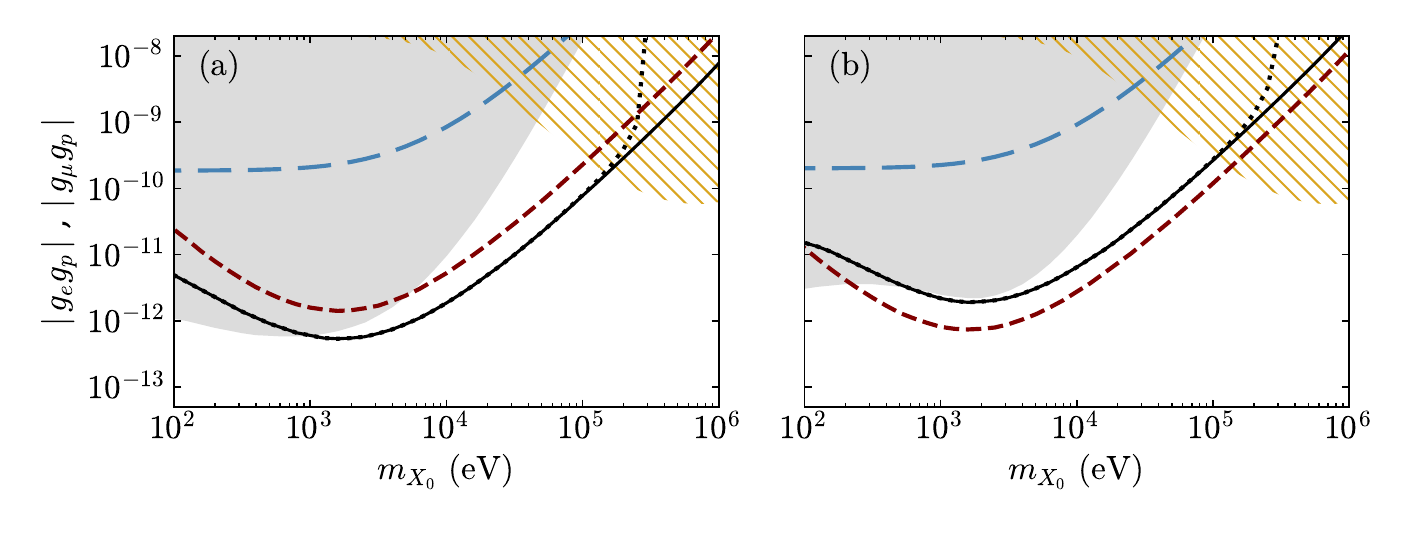}
\end{center}
\caption{Bounds on $g_eg_p$, (a) for an attractive NP interaction, (b) for a repulsive NP interaction. Shaded areas: regions excluded by the spectroscopy of eH \cite{Potvliege2023}. Solid curves: bounds based on the World spectroscopic data, assuming that $g_\mu = g_e$. Dotted curves: the same bounds as the solid curves but for the less constraining assumption that $-g_e\leq g_\mu \leq 100\, g_e$. Long-dashed curves: bounds based only on the 1s~--~2s interval of eH, the isotope shift of the 1s~--~2s interval and the $\mu$H and $\mu$D Lamb shifts, assuming that $g_\mu = g_e$. Short-dashed curves: the same as the long-dashed curves when the 2s~--~4p interval of eH is added to the data set. Hatched areas: values of $|g_\mu g_p|$ for which the new physics interaction between the muon and the proton would shift the 2s$_{1/2}$~---~2p$_{3/2}$ interval in muonic hydrogen by more than 5\% of the experimental error on the Lamb shift \cite{Potvliege2023}.}
\label{fig:gp}
\end{figure}

\begin{figure}
\begin{center}
\includegraphics[width=0.9\textwidth]{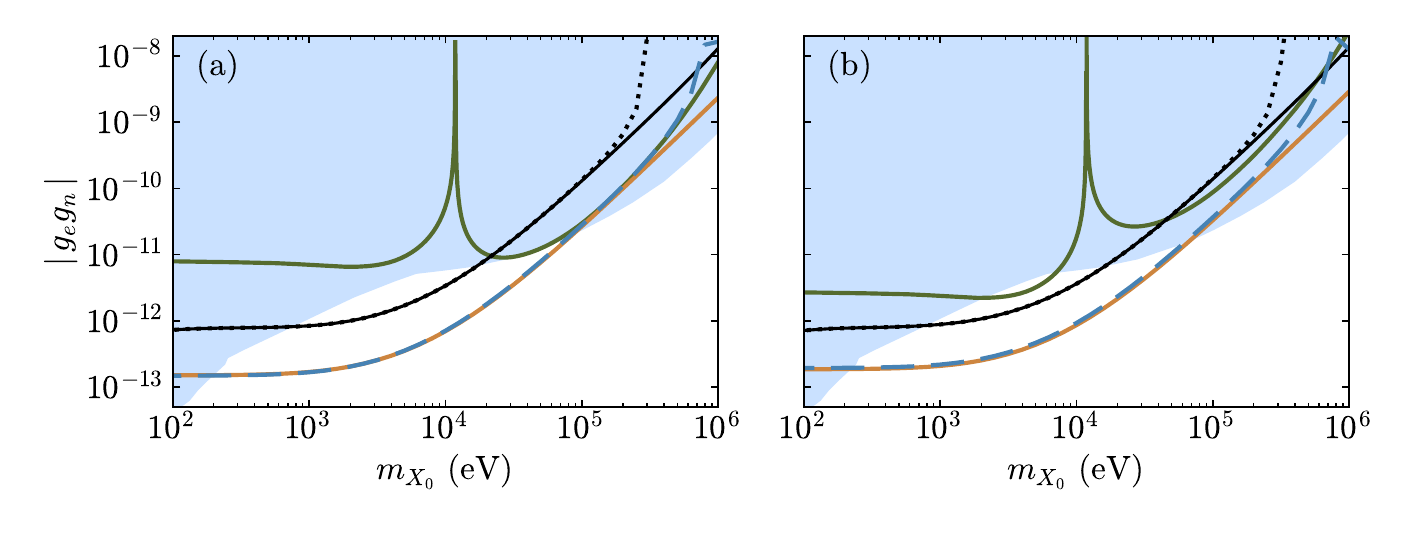}
\end{center}
\caption{Bounds on $g_eg_n$, (a) for an attractive NP interaction, (b) for a repulsive NP interaction. Shaded area: region excluded by neutron scattering data combined with measurements of the anomalous magnetic moment of the electron \cite{Berengut2018,Delaunay2017}. Solid green curves: upper bounds derived from the Yb/Yb$^+$ isotope shift \cite{Hur2022}. Solid black curves, dotted curves and long-dashed curves: as in Fig.~\ref{fig:gp}, here for $g_eg_n$. Solid orange curves, almost identical to the long-dashed curves below 100~keV: upper bounds calculated from Eq.~(\ref{eq:boundsiso}).}
\label{fig:gn}
\end{figure}

Proceeding as described in Section~\ref{section:general} yields bounds depending both on the value of the ratio $g_\mu/g_e$ and on the experimental data used in the calculation. We first consider results obtained under the lepton universality assumption that $g_\mu = g_e$. The bounds represented by the solid black curves in Figs.~\ref{fig:gp} and \ref{fig:gn} are based on the World spectroscopic data as in \cite{Potvliege2023}.\footnote{As in \cite{Potvliege2023,Tiesinga2021}, we alleviated difficulties in the $\chi^2$-fitting caused by the internal inconsistencies of this data set by magnifying all the experimental errors by 60\% when calculating these bounds and those represented by the dotted curves. The errors were not magnified for the other bounds discussed in this article.}
Previous results are also shown, for comparison. The shaded region in Fig.~\ref{fig:gp} identifies the values of $g_eg_p$ excluded by the spectroscopy of eH alone \cite{Potvliege2023}. The shaded region in Fig.~\ref{fig:gn} identifies the values of $g_eg_n$ excluded by an analysis of neutron scattering data and measurements of the anomalous magnetic moment of the electron \cite{Berengut2018,Delaunay2017}. The solid green curves, in Fig.~\ref{fig:gn}, show the bounds on $g_eg_n$ derived in \cite{Hur2022} from a generalized King plots analysis of the spectroscopy of Yb and Yb$^+$ (similar but slightly more constraining bounds have been obtained in still more recent analyses of King plots nonlinearities of isotope shifts of ytterbium and calcium transitions \cite{Door2024,Wilzewski2024}). 

Tighter bounds can be obtained by using a smaller set of experimental data in the fitting procedure. Only using the $\mu$H and $\mu$D data, the highly precise value of 1s$_{}$~--~2s$_{}$ transition frequency in eH \cite{Parthey2011} and the isotope shift of this interval \cite{Parthey2010} gives the tightest bounds on $g_eg_n$ that can be derived from hydrogen and deuterium spectroscopy \cite{Delaunay2017,Potvliege2023}. These bounds are represented by long-dashed curves in Fig.~\ref{fig:gn}.\footnote{These results are practically identical to those represented by the ``1S--2S HD (muonLS)'' curves in Figs.~1 and 3 of \cite{Delaunay2017}, which in effect were showing the $\pm 1.96\, \sigma_{g_eg_n}$ confidence interval defined in Section~\ref{section:sqraddiff} below rather than bounds as discussed here.} As seen from the graphs, they are significantly lower than the bounds based on the whole World data set and those based on King plots nonlinearities, except in the high mass region. 

The corresponding 1s~--~2s bounds on $g_eg_p$ are considerably weaker than those based on the World data (Fig.~\ref{fig:gp}), though, because a calculation based merely on that interval does not  strongly constrain $g_eg_p$ when $g_d \approx g_p$ \cite{Potvliege2023}. When $g_d = g_p$, indeed, the isotope shift depends on the strength of the new physics interaction only because of the difference in reduced mass between the two isotopes. This lack of sensitivity can be remedied by adding the transition frequency of a different interval to the data set, as long as the new physics shift of this interval differs enough from that of the 1s$_{}$~--~2s$_{}$ interval and the experimental error is sufficiently small. 
For example, and as seen from Fig.~\ref{fig:gp}, adding the 2s$_{}$~--~4p$_{}$ transition frequency measured in eH \cite{Beyer2017} yields bound comparable or tighter than those based on the World data. 

\subsection{Dependence on $g_\mu$}
\label{section:gmu}

How much the results of the previous section depend on the value of $g_\mu$ compared to the value of $g_e$ may be inferred from Appendix~A of \cite{Potvliege2023}, which concerns the impact of a new physics interaction on the determination of $r_p$ and $r_d$ from muonic hydrogen and muonic deuterium spectroscopy. The calculations described in that previous work aimed at delineating the values of $|g_\mu g_p|$ and $|g_\mu g_d|$ above which the interaction might affect the 2s$_{1/2}$~--~2p$_{3/2}$ interval significantly ($r_p$ and $r_d$ are derived from the experimental values of that interval). 
As in \cite{Potvliege2023}, we conservatively take ``significant" as meaning a shift of more than 5\% of the experimental error on the respective Lamb shift. The regions of the $(|g_\mu g_p|,m_{X_0})$ plane in which the impact of a new physics interaction is significant by that definition is represented by hashed areas in Fig.~\ref{fig:gp}.
It extends down to $|g_\mu g_p| \approx 2 \times 10^{-8}$ in the low mass region, and, as can be seen in the figure, to slightly below $1 \times 10^{-10}$ for $m_{X_0} \approx 1$~MeV. Up to 10~keV, $|g_\mu|$ should thus be at least four orders of magnitude larger than $|g_e|$ for invalidating the bounds discussed in the previous section. However, a smaller ratio of $|g_\mu|$ to $|g_e|$ would be sufficient to do so above 10~keV, particularly above 100~keV.

We examined the impact of a possible difference between $g_\mu$ and $g_e$ by recalculating the World data bounds under the more general assumption that $-g_e \leq g_\mu \leq 100g_e$. The calculation yield the bounds represented by the black dotted curves in Figs.~\ref{fig:gp} and \ref{fig:gn}. As expected, these bounds are practically identical to those obtained for $g_\mu = g_e$ for carrier masses below 100~keV. However, they are considerably less tight for higher masses.

\subsection{Bounds based on the difference $r^2_d - r^2_p$}
\label{section:sqraddiff}

As observed in Section~\ref{section:data}, there is excellent agreement in the value of $r^2_d - r^2_p$ between the result derived from the measurements in $\mu$H and $\mu$D and the result derived from the isotope shift of the 1s$_{}$~--~2s$_{}$ interval in eH and eD. Setting bounds on $g_eg_n$ based on these two results can be done as follows \cite{Delaunay2017}.
Let   
\begin{equation}
\delta r^2_{\mu,{\rm SM}} = r_d^2-r_p^2,
\end{equation}
where we take $r_d$ and $r_p$ to be the charge radii derived from the measurements in muonic deuterium and muonic hydrogen according to standard model theory: $\delta r^2_{\mu,{\rm SM}} = 3.8200(31)$~fm$^2$ \cite{Pachucki2024}. Similarly, let 
\begin{equation}
\delta r^2_{e,{\rm SM}} = r_d^2-r_p^2,
\end{equation}
here with the difference $r_d^2-r_p^2$ directly determined from the measurements of the isotope shift of the 1s$_{}$~--~2s$_{}$ interval in eD and eH, also according to standard model theory ($\delta r^2_{e,{\rm SM}} = 3.8207(3)$~fm$^2$ \cite{Pachucki2018}).
Also, let  
\begin{equation}
\Delta r^2_{2{\rm s}1{\rm s}} = \delta r^2_{\mu,{\rm SM}} - \delta r^{2}_{e,{\rm SM}}.
\end{equation}
Bounds on new physics may be sought in terms of the values of $g_eg_n$ for which $\Delta r^2_{2{\rm s}1{\rm s}}$ differs more from zero than would be expected in view of the experimental and theoretical errors on $\delta r^2_{\mu,{\rm SM}}$ and $\delta r^{2}_{e,{\rm SM}}$.  

The difference $\delta r^{2}_{e,{\rm SM}}$ is determined by equating the experimental isotope shift of the 1s$_{}$~--~2s$_{}$ interval, $\Delta \nu_{2{\rm s}1{\rm s}}^{\rm exp} = \nu_{2{\rm s}1{\rm s},{\rm eD}}^{\rm exp} - \nu_{2{\rm s}1{\rm s},{\rm eH}}^{\rm exp}$, to its standard model prediction,
$\Delta \nu_{2{\rm s}1{\rm s}}^{\rm SM} = \nu_{2{\rm s}1{\rm s},{\rm eD}}^{\rm SM} - \nu_{2{\rm s}1{\rm s},{\rm eH}}^{\rm SM}$. The latter can be separated into a term $\Delta \nu_{2{\rm s}1{\rm s}}^{\rm SM0}$ which does not depend sensitively on $r_p$ or $r_d$ and a term proportional to $\delta r^2_{e,{\rm SM}}$. Namely,
\begin{equation}
\Delta \nu_{2{\rm s}1{\rm s}}^{\rm SM} = \Delta \nu_{2{\rm s}1{\rm s}}^{\rm SM0} + {\cal C}_{2{\rm s}1{\rm s}}\delta r^2_{e,{\rm SM}},
\end{equation}
where ${\cal C}_{2{\rm s}1{\rm s}} = -1369.5$~kHz~fm$^{-2}$.
Thus
\begin{equation}
\delta r^2_{e,{\rm SM}} =  \frac{\Delta \nu_{2{\rm s}1{\rm s}}^{\rm exp} - \Delta \nu_{2{\rm s}1{\rm s}}^{\rm SM0}}{{\cal C}_{2{\rm s}1{\rm s}}}.
\label{eq:deltar2SM}
\end{equation}
Let us suppose that the experimental isotope shift $\Delta \nu_{2{\rm s}1{\rm s}}^{\rm exp}$ would differ from $\Delta \nu_{2{\rm s}1{\rm s}}^{\rm SM}$ by a new physics contribution $\Delta \nu_{2{\rm s}1{\rm s}}^{\rm NP}$, and let 
\begin{equation}
\delta r^2_e =  \frac{\Delta \nu_{2{\rm s}1{\rm s}}^{\rm exp} - \Delta \nu_{2{\rm s}1{\rm s}}^{\rm SM0}-\Delta \nu_{2{\rm s}1{\rm s}}^{\rm NP}}{{\cal C}_{2{\rm s}1{\rm s}}} = \delta r^2_{e,{\rm SM}} - \Delta \nu_{2{\rm s}1{\rm s}}^{\rm NP}/{\cal C}_{2{\rm s}1{\rm s}}.
\end{equation}
A non-zero $\Delta \nu_{2{\rm s}1{\rm s}}^{\rm NP}$ would make $\delta r^2_e$ a better approximation of the true value of $r_d^2-r_p^2$ than $\delta r^2_{e,{\rm SM}}$.
If we now assume that the new physics interaction does not significantly affect the measurements in the muonic species, then equating $\delta r^2_e$ to $\delta r^2_{\mu,{\rm NP}}$ gives
\begin{equation}
\Delta r^2_{2{\rm s}1{\rm s}} = \Delta \nu_{2{\rm s}1{\rm s}}^{\rm NP}/{\cal C}_{2{\rm s}1{\rm s}}.
\label{eq:Deltar2}
\end{equation}
The results presented in section~\ref{section:gmu} indicate that this assumption is unsafe for $m_{X_0} > 100$~keV. Accordingly, we do not consider this high mass region here. Below 100~keV, by contrast, these results indicate that a possible dependence of $r_d^2-r_p^2$ on $g_\mu$ seems unlikely in view of the lack of sensitivity of the measured $\mu$H and $\mu$D Lamb shifts on the value of this coupling constant in that mass region. Therefore, in common with \cite{Delaunay2017}, we take $g_\mu$ to be zero in the present approach. 

Apart from negligible differences in the wave functions arising from the different reduced masses, $\Delta \nu_{2{\rm s}1{\rm s}}^{\rm NP}$ is proportional to the difference $g_eg_d - g_eg_p$, which is $g_eg_n$.
We write 
\begin{equation}
\Delta \nu_{2{\rm s}1{\rm s}}^{\rm NP} = (g_eg_d - g_eg_p) \Delta\tilde{\nu}_{2{\rm s}1{\rm s}}^{\rm NP} = g_eg_n \Delta\tilde{\nu}_{2{\rm s}1{\rm s}}^{\rm NP}, 
\end{equation}
where $\Delta\tilde{\nu}_{2{\rm s}1{\rm s}}^{\rm NP}$ does not depend on $g_e$ or $g_n$. We also set\footnote{Apart from insignificant numerical differences beyond an overall factor of 1.96, the quantity $\sigma_{g_eg_n}$ defined by this equation is equivalent to the sensitivity parameter $\sigma(|g_eg_p|)$ defined by Eq.~(30) of \cite{Potvliege2023} if $g_d$ is taken to be $2 g_p$ when calculating the latter.}
\begin{equation}
\sigma_{g_eg_n} = \left|\frac{{\cal C}_{2{\rm s}1{\rm s}} \sigma_{2{\rm s}1{\rm s}}} {\Delta\tilde{\nu}_{2{\rm s}1{\rm s}}^{\rm NP}}\right|,
\label{eq:sigmagegnHD}
\end{equation}
where $\sigma_{2{\rm s}1{\rm s}}$ is the combined experimental and theoretical error on $\delta r^2_{\mu,{\rm SM}} - \delta r^{2}_{e,{\rm SM}}$. Assuming that $\Delta r^2_{2{\rm s}1{\rm s}}$ is not affected by systematic or random errors not taken into account through $\sigma$,
Eq.~(\ref{eq:Deltar2}) then implies that
\begin{equation}
\frac{{\cal C}_{2{\rm s}1{\rm s}} \Delta r^2_{2{\rm s}1{\rm s}}}{\Delta\tilde{\nu}_{2{\rm s}1{\rm s}}^{\rm NP}} - 1.96\,\sigma_{g_eg_n} 
\leq g_eg_n \leq \frac{{\cal C}_{2{\rm s}1{\rm s}} \Delta r^2_{2{\rm s}1{\rm s}}}{\Delta\tilde{\nu}_{2{\rm s}1{\rm s}}^{\rm NP}} + 1.96\,\sigma_{g_eg_n}
\label{eq:boundsiso}
\end{equation}
at the 95\% confidence level. The precision on the value of $g_eg_n$ obtained in this way is thus determined by $\sigma_{g_eg_n}$. This quantity can also be understood as quantifying the sensitivity of the method to a non-zero value of $g_eg_n$. 

The bounds given by Eq.~(\ref{eq:boundsiso}) are also plotted in Fig.~\ref{fig:gn}, where they are represented by the solid orange curves.
Up to masses of about 100~keV, these results are practically identical to the bounds represented by the long-dashed curves, which are based on the same experimental data but are obtained differently. The significant differences noticeable for higher masses illustrate the importance of allowing for a possible new physics contribution to the Lamb shift of the muonic species in that region.

\section{Extension to ${}^3\mbox{He}$ and ${}^4\mbox{He}$}
\label{section:3}

The approach to bounding $g_eg_n$ outlined in Section~\ref{section:sqraddiff} can be immediately extended to helium, now working with the difference $r_h^2 - r_\alpha^2$ between the squared nucleus rms charge radii of $^3$He and that of $^4$He rather than on the difference $r_d^2-r_p^2$ \cite{Delaunay2017}. This approach also avoids the need of taking into account a possible new physics interaction between the two electrons (see, e.g., \ref{section:appendixA}). As noted in Section~\ref{section:Hespectroscopy}, the necessary experimental isotope shifts are available for three different intervals, i.e., the 2$\,{}^3$S~--~2$\,{}^1$S and 2$\,{}^3$S~--~2$\,{}^3$P intervals, and, with a much lower precision, the 2$\,{}^1$S~--~3$\,{}^1$D interval.

Proceeding as in Section~\ref{section:sqraddiff} yields the 95\%-percent confidence bound
\begin{equation}
\frac{{\cal C}_{ba} \Delta r^2_{ba}}{-\Delta\tilde{\nu}_{ba}^{\rm NP}} - 1.96\,\sigma_{g_eg_n} 
\leq g_eg_n \leq \frac{{\cal C}_{ba} \Delta r^2_{ba}}{-\Delta\tilde{\nu}_{ba}^{\rm NP}} + 1.96\,\sigma_{g_eg_n}.
\label{eq:boundshe}
\end{equation}
Here $\Delta r^2_{ba}$ is the difference between the value of $r_h^2-r_\alpha^2$ derived from measurements in muonic helium ($\delta r^2_{\mu,{\rm SM}}$) and the value of this quantity derived from measurements of the isotope shift of the $a$~--~$b$ interval in electronic $^3$He and $^4$He according to standard model theory ($\delta r^2_{e,{\rm SM}}$). Moreover,
\begin{equation}
\sigma_{g_eg_n} = \left|\frac{{\cal C}_{ba} \sigma_{ba}}{\Delta\tilde{\nu}_{ba}^{\rm NP}}\right|,
\label{eq:sigmagegnHe}
\end{equation}
where $\sigma_{ba}$ is the combined experimental and theoretical error on $\Delta r^2_{ba}$. The minus signs in the denominators arise from the definition of the new physics contribution to the isotope shift, $\Delta \nu_{ba}^{\rm NP}$, which we take to be $\nu_{ba,^3{\rm He}}^{\rm NP} - \nu_{ba,^4{\rm He}}^{\rm NP}$ for consistency with the usual definition of the isotope shift for these intervals and the sign of ${\cal C}_{ba}$: here 
\begin{equation}
\Delta \nu_{ba}^{\rm NP} = (g_eg_h - g_eg_\alpha) \Delta\tilde{\nu}_{ba}^{\rm NP} = -g_eg_n \Delta\tilde{\nu}_{ba}^{\rm NP}.
\end{equation}
We use the values of ${\cal C}_{ba}$, $\Delta r^2_{ba}$ and $\sigma_{ba}$ listed in Table~\ref{table:table1} (we do not consider the 2$\,{}^1$S~--~3$\,{}^1$D interval in view of the large uncertainty on its isotope shift). The calculation of $\Delta\tilde{\nu}_{ba}^{\rm NP}$ is outlined in \ref{section:appendixA}.
\begin{table}
\label{table:table1}
\caption{
Data used in the present work. We assume that $\delta r^2_{\mu,{\rm SM}} = 3.8200(31)$~fm$^2$ for muonic hydrogen \cite{Pachucki2024}  and $1.0626(29)$~fm$^2$ for muonic helium \cite{LiMuli2024}.
For $\delta r^2_{e,{\rm SM}}$, we use the results of \cite{Pachucki2018} for the 1s$_{}$~--~2s$_{}$ interval,
the results of van der Werf {\it et al} \cite{Pachucki2024b,vanderWerf2023} for the 2$\,{}^3$S~--~2$\,{}^1$S interval, and the results of Shiner {\it et al} \cite{Pachucki2017,Shiner1995} and Cancio Pastor {\it et al} \cite{Pachucki2017,CancioPastor2004,CancioPastor2012} for the 2$\,{}^3$S~--~2$\,{}^3$P interval. The values of ${\cal C}_{ba}$ for helium are taken from \cite{Pachucki2017}.
}
\begin{indented}
\lineup
\item[]
\begin{tabular}{lcccc}
\br
Interval & $\delta r^2_{e,{\rm SM}}$~(fm$^2$) & $\Delta r^2_{ba}$~(fm$^2$) & $\sigma_{ba}$~(fm$^2$) & ${\cal C}_{ba}$~(kHz~fm$^{-2}$) \\ 
\mr
1s$_{}$~--~2s$_{}$ & 3.8207(3)  & $-0.0007$  & $0.0031$  & $-1369.5\0\0$ \\
2$\,{}^3$S~--~2$\,{}^1$S & $1.0678(7)$ & $-0.0052$ & $0.0030$ & $-214.66$  \\
2$\,{}^3$S~--~2$\,{}^3$P (S) & $1.061(3)\0$ & \m $0.002\0$ & $0.004\0$  & $-1212.2\0\0$\\
2$\,{}^3$S~--~2$\,{}^3$P (CP) & $1.069(3)\0$ & $-0.006\0$ & $0.004\0$ & $-1212.2\0\0$\\
\br
\end{tabular}
\end{indented}
\end{table}

\begin{figure}
\begin{center}
\includegraphics[width=0.6\textwidth]{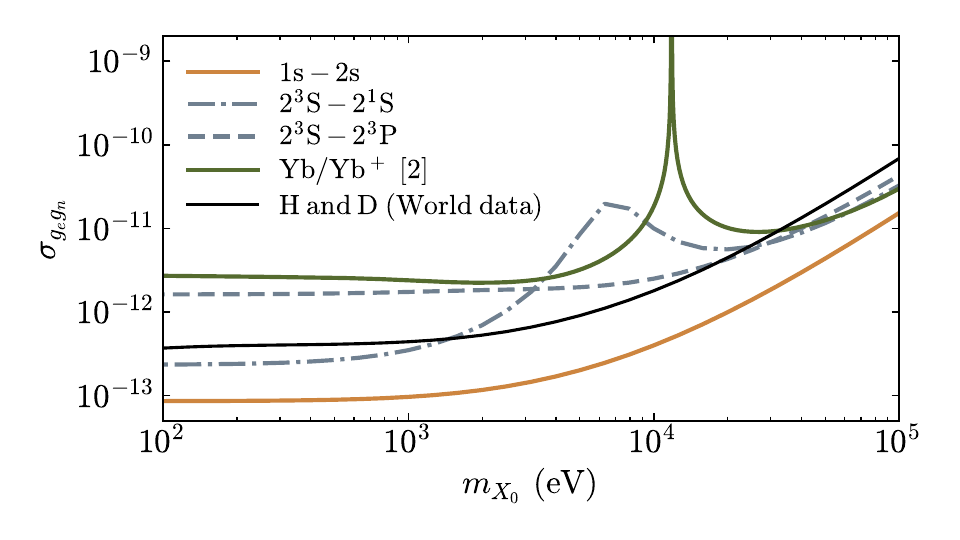}
\end{center}
\caption{The sensitivity parameter $\sigma_{g_eg_n}$ of Eqs.~(\ref{eq:sigmagegnHD}) and (\ref{eq:sigmagegnHe}) as derived from the isotope shift of the 1s~--~2s interval in hydrogen (solid orange curve), from the isotope shift of the $2^3\mbox{S}$~--~$2^1\mbox{S}$ interval of He (dash-dotted curve) or from the isotope shift of the $2^3\mbox{S}$~--~$2^3\mbox{P}$ interval of He (short-dashed curve). The solid green curve and solid black curve indicate the corresponding values of $\sigma_{g_eg_n}$ for, respectively, the King plots analysis of \cite{Hur2022} and the World data results of Section~\ref{section:results}, as explained in the text.}
\label{fig:He_sigma}
\end{figure}
The resulting values of $\sigma_{g_eg_n}$ are plotted in Fig.~\ref{fig:He_sigma}, both for helium and for the 1s$_{}$~--~2s$_{}$ interval of hydrogen.
The form of $V_{\rm NP}(r)$ implies that in the $m_{X_0}\rightarrow 0$ limit
\begin{equation}
\sigma_{g_eg_n} \propto 
\left|
\frac{{\cal C}_{ba}\sigma_{ba}}{n\langle b | 1/r | b \rangle - n\langle a | 1/r | a \rangle}
\right|,
\label{eq:sensitivity2}
\end{equation}
where $n$ is the number of electrons. ${\cal C}_{ba}$ is larger for the 1s$_{}$~--~2s$_{}$ interval of hydrogen than for the two intervals of helium considered here. However, the denominator of Eq.~(\ref{eq:sensitivity2}) is considerably smaller for the latter \cite{Davis1982}, with the consequence that a greater sensitivity is obtained in the low mass region by using the 1s$_{}$~--~2s$_{}$ interval. As can be seen from the figure, this is also the case beyond that region. Bounds based on the isotope shift of these two helium intervals can therefore be expected to be less tight than the bounds based on the 1s$_{}$~--~2s$_{}$ interval.

The sensitivity to a non-zero value of $g_eg_n$ of the King plots analysis of \cite{Hur2022} and of the World data results of Section~\ref{section:results} is also indicated in Fig.~\ref{fig:He_sigma}. For these two approaches, we take $\sigma_{g_eg_n}$ to be half the width of the respective 95\% confidence interval on the value of $g_eg_n$, divided by 1.96 for consistency with Eqs.~(\ref{eq:sigmagegnHD}) and (\ref{eq:sigmagegnHe}). More recent results based on ytterbium and calcium spectroscopy \cite{Door2024,Wilzewski2024} have improved the sensitivity of the methods based on King plots analyses by roughly a factor of two. However, at the present time and within the current state of development of these methods, the comparison points to a greater sensitivity of the spectroscopy of hydrogen and (to a lesser extent) of helium in regard to the detection of a fifth force.

\begin{figure}
\begin{center}
\includegraphics[width=0.9\textwidth]{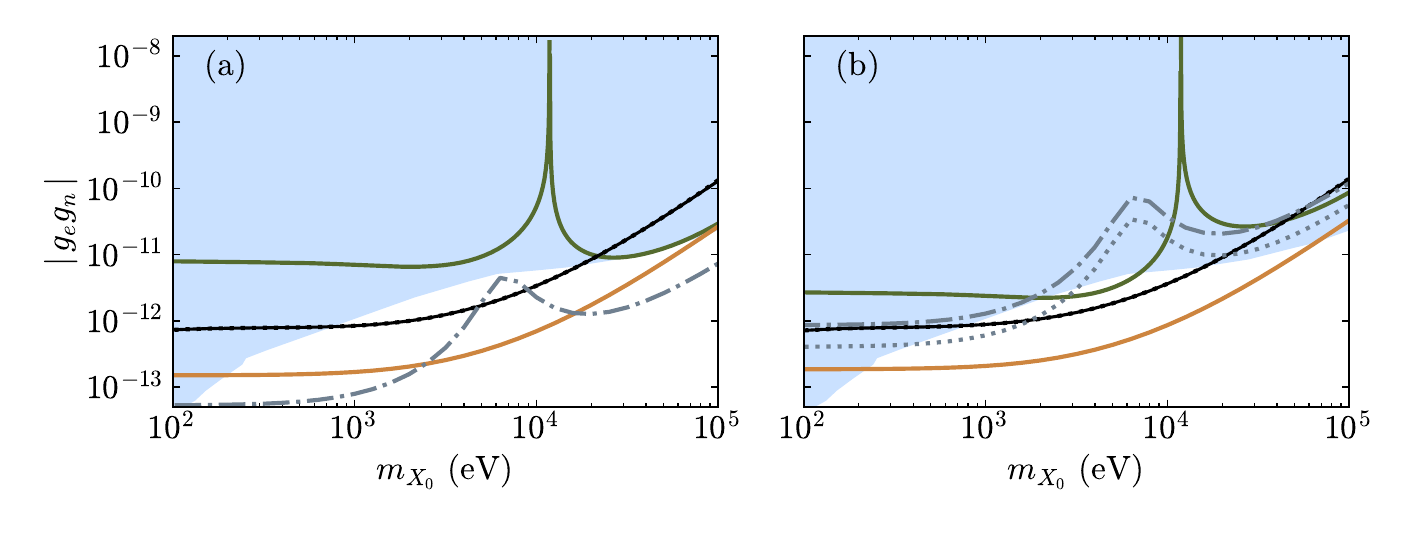}
\end{center}
\caption{As in Fig.~\ref{fig:gn}, here with dash-dotted curves indicating the bounds on $g_eg_n$ based on the isotope shift of the $2^3\mbox{S}$~--~$2^1\mbox{S}$ interval of He. The dotted curve in (b) is the value of $g_eg_n$ for which the new physics shift would entirely explain the difference between $\delta r^2_{\mu,{\rm SM}}$ and $\delta r^2_{e,{\rm SM}}$.}
\label{fig:gn_He}
\end{figure}
The bounds on the value of $g_eg_n$ predicted by Eq.~(\ref{eq:boundshe}) for the $2^3\mbox{S}$~--~$2^1\mbox{S}$ interval are shown in Fig.~\ref{fig:gn_He}, where they are represented by dash-dotted curves (the corresponding results for the $2^3\mbox{S}$~--~$2^3\mbox{P}$ interval are presented in the Supplementary Material, for completeness). These bounds are not symmetrical around $g_eg_n =0$ because of the significant difference between $\delta r^2_{\mu,{\rm SM}}$ and $\delta r^2_{e,{\rm SM}}$ for this interval, with the result that the $2^3\mbox{S}$~--~$2^1\mbox{S}$ bound tends to be particularly tight in Fig.~\ref{fig:gn_He}(a). However, this difference seems too large to be primarily due to a new physics shift, if there would be any suspicion that a fifth force might be at play here, as can be surmised from the dotted curve indicating the centre of the confidence interval defined by Eq.~(\ref{eq:sensitivity2}).
The values of $g_eg_n$ represented by this curve are larger than the corresponding 1s~--~2s bound (the solid orange curve), and are therefore excluded by it. Taking the $2^3\mbox{S}$~--~$2^1\mbox{S}$ bound of Fig.~\ref{fig:gn_He}(a) at face value would therefore be imprudent. Nonetheless, it is worth noting that these helium results are broadly consistent with the bounds derived from hydrogen and deuterium spectroscopy.

\section{Conclusions}

In summary, we have presented newly calculated bounds on the products $g_eg_p$ and $g_eg_n$ derived from the high precision spectroscopic data currently available for hydrogen, deuterium, helium-3 and helium-4. These results update those of \cite{Delaunay2017} and build up on our previous work on the topic \cite{Jones2020,Potvliege2023}. They do not depend on a specific assumption on the ratio $g_d/g_p$ (or on the ratio $g_\alpha/g_h$ in helium), contrary to the confidence intervals on $g_eg_p$ presented in \cite{Potvliege2023}. They do depend on the ratio $g_\mu/g_e$, but in a minor if not completely negligible way for carrier masses below 100~keV if $|g_\mu|$ is assumed not to be several order of magnitude larger than $|g_e|$. 

In this mass region, the bounds on $g_eg_n$ based on the World spectroscopic data for hydrogen and deuterium tend to be more stringent than the bounds arising from the analysis of King plots nonlinearities, in the current state of development of that approach \cite{Hur2022,Door2024,Wilzewski2024}. However, they are impacted by the well known inconsistencies between the available data. As was already pointed out in \cite{Delaunay2017}, particularly stringent bounds can be set by combining the isotope shift of the 1s~--~2s interval in eH and eD with the rms charge radii of the proton and the deuteron derived from the measurements of the Lamb shift in $\mu$H and $\mu$D. However, being based on a relatively small number of experiments, these results might conceivably be affected by unknown systematic errors.

Setting bounds based on the isotope shift of particular intervals in helium is also possible. Doing so results in bounds broadly consistent with those obtained for hydrogen and deuterium. The approach is less powerful for helium, though, because of the smaller new physics shift of the intervals for which sufficiently precise isotope shifts are available.

The theoretical error on the value of $r_d^2-r_p^2$ derived from muonic hydrogen and muonic deuterium is the main limitation on the sensitivity of the bounds based on the isotope shift of the 1s~--~2s interval in hydrogen and deuterium. Lowering this theoretical error would thus make it possible to strengthen these bounds further. Alternatively, the same could also be achieved by combining the isotope shift of this interval with that of another interval, which would also bypass the need of using the muonic species data and therefore eliminate the dependence of these bounds on the value of $g_\mu$ \cite{Delaunay2017,Potvliege2023}.

\section*{Acknowledgements}
This work much benefitted from conversations with M P A Jones and M Spannowsky and communications from K Pachucki. The programs used in the helium calculations are based on codes written by B Yang, M Pont, T Li and R Shakeshaft and kindly provided to the author by R Shakeshaft a number of years ago.
\appendix

\section{Calculation of the new physics frequency shifts}
\label{section:appendixA}
A new physics interaction of the type considered in this work would  potentially affect the experimental transition frequencies by shifting the energies of the respective states from their standard model values. The contribution $\nu_{ba}^{\rm NP}$ this interaction would make to the transition frequency of a transition between a state $a$ and a state $b$ would be
\begin{equation}
\nu_{ba}^{\rm NP} = (\delta E_{b}^{\rm NP} - \delta E_{a}^{\rm NP})/h,
\end{equation}
in terms of the new physics shifts $\delta E_{a}^{\rm NP}$ and
$\delta E_{b}^{\rm NP}$ of the energies of the respective states and of the Planck's constant $h$.
Since the potential $V_{\rm NP}(r)$ is certainly very weak compared to the Coulomb potential, if non-zero, the energy shifts $\delta E_{a}^{\rm NP}$, $\delta E_{b}^{\rm NP}$, \ldots, do not need to be calculated beyond first order perturbation theory \cite{Jaeckel2010,Karshenboim2010b,Brax2011}. 

Accordingly, we simply set, for electronic hydrogen and deuterium,
\begin{align}
\delta E_{a}^{\rm NP} =
\int \psi_{a}^*({\bf r}) V_{\rm NP}(r) 
\psi_a({\bf r})\,{\rm d}{\bf r},\qquad
\delta E_{b}^{\rm NP} =
\int \psi_{b}^*({\bf r}) V_{\rm NP}(r) 
\psi_b({\bf r})\,{\rm d}{\bf r},
\end{align}
where $\psi_a({\bf r})$ and $\psi_b(\bf r)$ are the unperturbed non-relativistic wave functions of the corresponding bound states. As in \cite{Jones2020,Potvliege2023}, we calculate these energy shifts either analytically or numerically, in the latter case by obtaining the wave functions by diagonalising the matrix representing the unperturbed Hamiltonian in a Sturmian basis. For muonic hydrogen and muonic deuterium, we use relativistic wave function obtained by solving the Dirac equation for a muon in the Coulomb and Uehling potentials of an extended nuclear charge distribution, as in \cite{Potvliege2023}.

A similar calculation for helium would normally involve computing matrix elements of the new physics electron-electron interaction, besides computing the matrix elements of $V_{\rm NP}(r)$ for each of the electrons. A method for doing this is described in \cite{Delaunay2017}. In the present work, however, we only consider the effect of the new physics interaction on the isotope shift of transitions in electronic $^3$He and $^4$He. We only need to calculate $\Delta \nu_{ba}^{\rm NP}$ for the relevant transitions, thus, rather than individual new physics energy shifts. In terms of the latter,
\begin{equation}
\Delta \nu_{ba}^{\rm NP} = [(\delta E_{b,^3{\rm He}}-\delta E_{a,^3{\rm He}})-(\delta E_{b,^4{\rm He}}-\delta E_{a,^4{\rm He}})]/h.
\end{equation}
At the level of the Schr\"odinger equation,
the contribution of the new physics electron-electron interaction to the energy shift of a same state differs between $^3$He and $^4$He only because of the different nuclear masses of these two isotopes, which impact on the wave functions through reduced mass and mass polarisation corrections. These differences are negligible for our purpose. Hence, only the electron-nucleus new physics interaction needs to be taken into account in the isotope shifts. In this approximation, and assuming that $g_\alpha - g_h = g_n$ as noted above,
\begin{align}
\delta E_{a,^4{\rm He}}-\delta E_{a,^3{\rm He}} =
(-1)^{s+1} g_eg_n 
\int \psi_{a}^*({\bf r}_1,{\bf r}_2) \left[\tilde{V}_{\rm NP}(r_1) + \tilde{V}_{\rm NP}(r_2)\right]  
\psi_a({\bf r}_1,{\bf r}_2)\,{\rm d}{\bf r}_1\,{\rm d}{\bf r}_1,
\end{align}
and similarly for the difference $\delta E_{b,^4{\rm He}}-\delta E_{b,^3{\rm He}}$. In this equation, ${\bf r}_1$ and ${\bf r}_2$ are the position vectors of the two electrons, $\psi_a({\bf r}_1,{\bf r}_2)$ is the unperturbed non-relativistic wave function of state $a$ for an infinite nuclear mass, and
\begin{equation}
\tilde{V}_{\rm NP}(r) = \frac{1}{4 \pi}~\frac{1}{r}~\exp(-m_{X_0} r).
\label{eq:yukpot3}
\end{equation}
An accurate calculation requires correlated two-electron wave functions \cite{Delaunay2017}. We use wave functions obtained by diagonalising the unperturbed Hamiltonian in a Laguerre basis expressed in perimetric coordinates, following \cite{Pekeris1958} and more recently \cite{Yang1997,Li2005}. Specifically, we use a basis of antisymmetrized products of an angular factor and radial functions of the following form
\cite{Yang1997,Li2005},
\begin{equation}
\phi_{lmn}(u,v,w) = \exp[-(k_1 u + k_2 v + k_3 w)/2]L_l(k_1 u)L_m(k_2 v) L_n(k_3 w),
\end{equation}
where $L_p(\cdot)$ denotes the Laguerre polynomial of order $p$, $k_1$ and $k_2$ are two scaling constants, $k_3 = (k_1 + k_2)/2$, and
\begin{equation}
u = r_2 + r_3 - r_1, \qquad v = r_1 + r_3 - r_2, \qquad w = 2(r_1 + r_2 - r_3)
\end{equation}
with $r_3 = |{\bf r}_1 - {\bf r}_2|$. We set $k_1 = 2.01 \,a_0^{-1}$ and $k_2 = 0.765\, a_0^{-1}$, where $a_0$ is the Bohr radius. This choice, while presumably not optimal, ensured that the expectation value of the new physics potential converged to between four and seven significant figures, depending on the state and on $m_{X_0}$, when the basis was increased to the maximum size used in the computation ($l + m + n \leq 10$ for the $2^1\mbox{S}$ state, $\leq 8$ for the $2^3\mbox{S}$ state and $\leq 15$ for the $2^3\mbox{P}$ state). These parameters also ensured that the corresponding values of $\langle 1/r \rangle$ matched the benchmark results of \cite{Davis1982} to five significant figures.

\end{document}